\documentstyle[twoside,fleqn,espcrc2,epsfig,rotate]{article}

\newcommand{\AmS}{{\protect\the\textfont2
  A\kern-.1667em\lower.5ex\hbox{M}\kern-.125emS}}

\title{Lattice Approach to Diquark Condensation in Dense Matter }

\author{Susan Morrison (UKQCD collaboration) \address{Department of Physics, 
        University of Wales Swansea, \\ 
        Singleton Park, Swansea SA2 8PP, U.K.}%
        \thanks{Supported in part by E.U. grant ERBFMRX-CT97-0122}}

\begin{document}

\begin{abstract}
Results are presented of a Monte Carlo simulation of a three dimensional 
Gross-Neveu
model with $\mbox{SU(2)}\otimes\mbox{SU(2)}$ chiral symmetry at non-zero baryon
chemical potential $\mu$, corresponding to non-zero baryon density. The 
phenomenon of quark pair condensation is sought via measurement of a 
two point function. For $\mu$ sufficiently large there is a sharp
transition between a phase where the chiral symmetry is broken by a 
condensate $\langle\bar qq\rangle$ and one where 
a scalar diquark condensate $\vert\langle qq\rangle\vert\not=0$.
Global U(1) baryon number
symmetry may remain unbroken, however, due to the absence of long range
order in the phase of $\langle qq\rangle$. 
\end{abstract}

\maketitle

Recent work \cite{ARW,RSSV} suggests that new 
forms of ordering may exist at high baryon density. In addition
to the standard $\langle \bar qq\rangle$ condensate associated with spontaneous 
breaking of chiral symmetry it has been proposed that the existence of a Fermi 
surface combined with any attractive $qq$ interaction results in the 
formation of a $\langle qq \rangle$ condensate of quark Cooper pairs. Remarkably,
this {\sl diquark condensate\/} is associated with baryon number violation. 
Formation of a diquark condensate, which is non-invariant
under the vectorlike U(1) global symmetry of baryon number is generally 
forbidden by the Vafa-Witten (VW) theorem \cite{VW}. Notable exceptions to the 
VW theorem include theories with complex measures and models with Yukawa couplings 
to scalar degrees of freedom, such as the coupling to $\sigma$ in (\ref{eq:Z})
below.

For $\mu\not= 0$ the fermion determinant of QCD is complex and lattice simulations  
fail dramatically however the lattice approach 
{\sl can \/} be successfully applied at $\mu\not=0$ to toy field theories.
I will present work done in collaboration with S.J. Hands which is the first lattice study of diquark condensation at $\mu \not=0$
implemented in the Gross-Neveu (GN) model with $\mbox{SU(2)}\otimes\mbox{SU(2)}$ chiral 
symmetry and formulated in $d=2+1$ spacetime dimensions. With the exception
of confinement this model incorporates all of the essential physics.

The lattice model we have simulated has the following Euclidean path
integral:
\begin{equation}
Z=\int D\sigma D\vec\pi\mbox{det}(M^\dagger M[\sigma,\vec\pi])e^{
-\frac{2}{g^2}(\sigma^2+\vec\pi.\vec\pi)}
\label{eq:Z}
\end{equation}
where $M$ is the staggered fermion matrix; $\sigma$ and the triplet 
$\vec\pi$ are real auxiliary
fields defined on the dual sites $\tilde x$
of a three dimensional Euclidean lattice. The lattice model is formulated such 
that the integration measure is real and positive. The full lattice action 
(involving isodoublet staggered lattice fermion fields $\chi$ and $\zeta$) and
its symmetries are fully discussed in \cite{HM98}. The corresponding
$d=3+1$ model is outlined in \cite{HK}.

 In the limit $m\to0$, our model is invariant under a global symmetry
akin to the $\mbox{SU(2)}_L\otimes\mbox{SU(2)}_R$ of the continuum 
NJL model. For $T=\mu=0$, a mean field treatment
shows that this $\mbox{SU(2)}\otimes\mbox{SU(2)}$ symmetry is
spontaneously broken to $\mbox{SU(2)}_{isospin}$ by the generation of a
condensate $\Sigma_0
=\langle\sigma\rangle={2\over g^2}\langle\bar\chi\chi\rangle$
and the fermion acquires physical mass $m_f=\Sigma_0$.The symmetry is broken for
coupling $g^2>g_c^2\simeq1.0$. 
The numerical results presented here are for a value 
$g^2=2.0$, corresponding to a theory deep in the broken phase for $\mu=0.0$,
and with $\Sigma_0=0.706(1)$ in units of inverse lattice spacing, ie.
rather far from the continuum limit.

For $T,\mu\not=0$ a mean field solution is also known \cite{HKK},
in which $\Sigma(\mu,T)$ is expressed in terms of $\Sigma_0$. At $T=0$
the basic feature is that $\Sigma$ remains constant as $\mu$ is 
increased up to a critical value $\mu_c=\Sigma_0$, whereupon $\Sigma$ falls
sharply to zero (in the chiral limit), signalling a first order chiral symmetry
restoring transition. Monte Carlo simulations \cite{HKK} support
this picture, even when massless Goldstone excitations
(with the quantum numbers of the $\pi$ field) are present \cite{HKK2}.

We performed simulations of the model (\ref{eq:Z})
on $L_s^2\times L_t$ lattices with bare mass $m=0.01$ at a coupling 
$1/g^2=0.5$. The simulation method is a standard
hybrid Monte Carlo algorithm. We monitored the 
chiral condensate $\langle\bar\chi\chi\rangle$, and the baryon density 
$n$ using stochastic estimators.

We have constructed two local diquark operators which are anti-symmetric in
spatial, flavour and isospin indices. Firstly the  
{\sl spectral scalar\/}:\,$\chi^p(x)\tau_2^{pq}\chi^q(x)$
which is even under parity and secondly the  {\sl spectral pseudoscalar\/}: 
$(-)^{x_1+x_2+x_3}\chi^p(x)\tau_2^{pq}\chi^q(x)$
which is odd under parity.

Measurement of diquark condensates on the lattice demands
innovative techniques and for this initial study we chose an indirect 
method: the two
point function $\langle qq(0)\bar q\bar q(x)\rangle$ 
was measured as the expectation
of the product of two fermion propagators as in QCD baryon spectroscopy, and
the condensate extracted by assuming clustering at large spacetime separation, 
ie.
\begin{equation}
\langle qq(0)\bar q\bar q(x)\rangle=
\langle qq(0)\bar q\bar q(x)\rangle_c+\langle qq\rangle\langle\bar q\bar
q\rangle,
\end{equation}
the latter term on the right being proportional to 
$\vert\langle qq\rangle\vert^2$. A non-zero condensate signal
therefore shows up as
a plateau in the timeslice correlator $G(t)$ at large time separation, and
may be extracted by a fit of the form
\begin{eqnarray}
G(t)&=&\sum_{x_1,x_2}\langle qq(0)\bar q\bar q(t,x_1,x_2)\rangle\nonumber\\
&=&A\exp(-M_+t)+B\exp(-M_-(L_t-t))\nonumber\\
& &+C^2(L_s)\vert\langle qq\rangle\vert^2,
\label{eq:timeslice}
\end{eqnarray}
with $M_\pm$ the masses of forward and backward propagating diquark states 
respectively. The value of the constant $C(L_s)$ is not determined {\sl a
priori\/}. Naively one expects $C=L_s$, and hence the plateau height to 
be an extensive quantity.
\begin{figure}
\epsfig{file=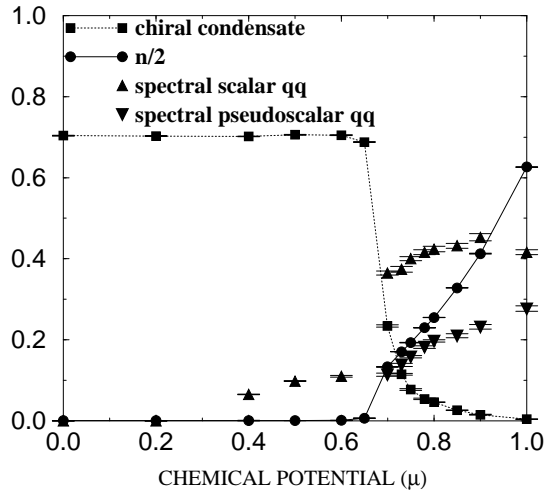,width=7.5cm,angle=0,clip=}
\vspace{-1.2cm}
\caption{Overview of observables $\langle\bar\chi\chi\rangle$, $n$ and
diquark condensate signals
$C\vert\langle qq\rangle\vert$ as functions of $\mu$, obtained on a 
$16^2\times24$ lattice, showing the onset
of diquark condensation at $\mu_c\simeq0.65$.}
\vspace{-0.8cm}
\label{fig:chir_transition}
\end{figure}
The trend in the diquark condensates as we pass from broken to symmetric phase
is clear from Fig. \ref{fig:chir_transition}. There is a critical value,
$\mu_c\simeq 0.65$, at which the chiral $\langle\bar{q}q\rangle$ condensate
falls sharply and the number density (n) begins to rise from zero. 
The spectral scalar 
$C\vert\langle qq\rangle\vert$ 
condensate rises slowly from zero for $0.4 \le \mu<\mu_c$,
jumps upwards discontinuously close to $\mu_c$, and
continues to rise steadily. For
$\mu>1.0$ as the number density approached saturation the scalar 
$qq$ propagators became distorted. 
The spectral
pseudoscalar $C\vert\langle qq \rangle\vert$ signal was consistent with zero
for $\mu<\mu_c$ and no fits to the propagators were possible but for
$\mu>\mu_c$ this condensate is non-zero and increases with $\mu$. It is 
considerably smaller in magnitude than the spectral scalar. This pseudoscalar 
condensate is {\sl parity violating}, and therefore it would be
remarkable if the signal in the chirally symmetric phase were to remain non-zero 
in the large volume limit.

The trend in the $M_\pm$ data \cite{HM98} was clear with small masses in 
the symmetric phase and large masses in the broken phase. The forward
propagating $M_+$ states were more difficult to
extract from the fits than the backward propagating $M_-$ states. 
For $\mu<\mu_c$, $M_-$ decreased linearly with $\mu$ reflecting the 
trend observed in the physical fermion mass $m_f$ in previous
simulations \cite{HKK2}.
$M_-$ was approximately constant for $\mu>\mu_c$. 
The zero density fermion mass $m_f=\Sigma_0\simeq0.7$,
defines the ratio of physical to cutoff scales and the observed 
$qq$ states had masses around 0.3 which is comparatively light.

Fits to the spectral scalar correlator 
in the symmetric phase ($\mu=0.8$) were very stable as $L_t$ was increased
from 16 to 40. The smaller parity violating signal, on the other hand, 
decreased  with increasing $L_t$. The non-zero spectral scalar
condensate in the broken phase is at least in part due to finite
lattice size.
\begin{figure}
\epsfig{file=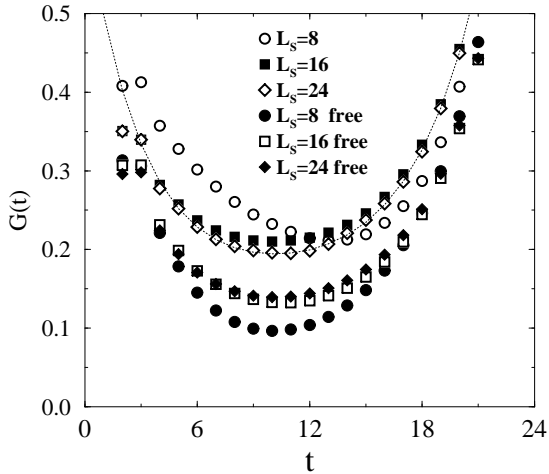,width=7.5cm,clip=}
\vspace{-1.2cm}
\caption{Spectral scalar diquark correlator 
at $\mu=0.8$ for both free and interacting quarks on $L_s^2\times24$ lattices.}
\vspace{-0.8cm}
\label{fig:volume}
\end{figure}
To determine the behaviour with spatial volume we next performed a series of
runs on $L_s^2\times24$ lattices at $\mu=0.8$, with $L_s$ varying from 8 to
24. To our surprise we found very little change as $L_s$ increased, the
fitted value of $C\vert\langle qq\rangle\vert$ 
more or less saturating for $L_s\geq16$.
Results for $L_s=8,16,24$ are shown in Fig. \ref{fig:volume}, together with
the fit for $L_s=24$. For comparison we also plot
the equivalent correlators for free fermions at $\mu=0.8$; it is striking that
for this case the trend as $L_s$ increases is in the opposite 
direction. We conclude that the constant $C$ is roughly independent
of $L_s$

The fact that the two-point correlation function exhibits clustering at 
large temporal separation is suggestive of diquark condensation
at $\mu\not= 0$.
In order to quantify the measurement, however, it will be necessary to provide a
numerical estimate for
the constant $C$ in (\ref{eq:timeslice}). 

The spontaneous breaking of the U(1) symmetry 
usually implies the existence of a Goldstone mode associated with long
wavelength fluctuations
in the phase of the $\langle qq\rangle$ condensate.
However, in the absence of an explicit diquark source (analogous
to a bare mass for the case of $\langle\bar qq\rangle$), one would expect
large finite volume effects, generically proportional to $L^{-(d^\prime-2)}$,
where $d^\prime$ is the dimension of the effective field theory
describing the fluctuations of the order parameter
\cite{Has}. We therefore speculate that the effective 
field theory describing the spatial correlations of the diquark
correlator has $d^\prime=d-1=2$, 
and that there is no long range order 
in the spatial directions \cite{Mermin}. A similar distinction between
temporal and spatial correlations has been observed in instanton liquid models
\cite{RSSV}.
This would account for the volume-independence of
$C$. Strictly speaking, therefore, we have not observed a true 
condensation. This interpretation of our results 
can be tested by simulations of the equivalent
$3+1$ dimensional model \cite{HK}, where we {\sl would\/} expect long range order.
An extension of this work will
involve the introduction of a diquark source term to study the
one point function. This will serve to calibrate our 
two point function measurement and to extend our study to other
diquark operators and the $T\not=0$ regime.

\end{document}